\documentstyle[twoside,fleqn,espcrc2,epsf]{article}

\newcommand{\AmS}{{\protect\the\textfont2
  A\kern-.1667em\lower.5ex\hbox{M}\kern-.125emS}}

\newcommand{\bd}{\begin{displaymath}}
\newcommand{\ed}{\end{displaymath}}
\newcommand{\be}{\begin{equation}}
\newcommand{\ee}{\end{equation}}
\newcommand{\bea}{\begin{eqnarray}} 
\newcommand{\eea}{\end{eqnarray}}
\newcommand{\bt}{\begin{tabular}}
\newcommand{\et}{\end{tabular}\newline}
\newcommand{\x}{\mbox{x}}

\title{
\vspace{-4.05cm}                                            
{\normalsize DESY 98--132}    \\[-0.2cm]                    
{\normalsize HUB--EP--98/60}  \\[-0.2cm]                    
{\normalsize TPR--98--28}     \\[-0.2cm]                    
{\normalsize September 1998}  \\                            
\vspace{2.25cm}                                             
Renormalization of four-fermion operators for higher twist 
calculations\thanks{Talk 
given by S. Capitani at Lattice 98,                         
               Boulder (US).}}                              

\author{
S.~Capitani\address{Deutsches Elektronen-Synchrotron DESY, Notkestrasse 85, 
D-22607 Hamburg, Germany}, 
M.~G\"ockeler\address{Universit\"at Regensburg, Institut f\"ur Theoretische 
Physik, D-93040 Regensburg, Germany}, 
R.~Horsley\address{Humboldt-Universit\"at, Institut f\"ur Physik, 
Invalidenstrasse 110, D-10115 Berlin, Germany}, 
H.~Perlt\address{Universit\"at Leipzig, Institut f\"ur Theoretische Physik,
Augustusplatz 10/11, D-04109 Leipzig, Germany}, 
P.~Rakow$^{\rm b}$, 
G.~Schierholz$^{\rm a,}$\address{Deutsches Elektronen-Synchrotron DESY and 
NIC, Platanenallee 6, D-15738 Zeuthen, Germany} 
and A.~Schiller$^{\rm d}$
}

\begin{document}

\begin{abstract}
The evaluation of the higher twist contributions to 
Deep Inelastic Scattering amplitudes involves a non trivial
choice of operator bases for the higher orders of the 
OPE expansion of the two hadronic currents. In this talk we
discuss the perturbative renormalization of the 
four-fermion operators that appear in the above bases.
\end{abstract}

\maketitle

\section{INTRODUCTION}

We present here the first steps of a more general and large scale research 
program, which aims at the evaluation of the contributions of higher twist 
operators in Deep Inelastic Scattering (DIS) processes~\cite{ht,js,lww,sv}. 
We consider in particular the next-to-leading terms in the OPE, which are the 
next order corrections to the presently known DIS amplitudes and structure 
functions.
The matrix elements of higher twist operators are outside the reach of 
perturbative QCD, and lattice QCD provides at present the only way to compute 
them in a reliable way and from first principles. Renormalization factors are 
then necessary to relate the numbers extracted directly from the lattice to 
the physical matrix elements.

\section{HIGHER TWIST}

In the light-cone expansion of the T-product of two hadronic currents in DIS, 
the leading behavior of the Wilson coefficients is given (up to logs) by 
$C^{n,i}(x^{2}) \sim (x^2)^{(d_{O^{(n,i)}} - n - 2d_J) / 2}$, 
and is governed by the twist $\tau = d_{O^{(n,i)}} - n$ 
(dimension minus spin) of the corresponding operators. The various 
terms in the expansion can thus be classified according to their twist:
operators of twist-2 are the leading ones, and the higher twist contributions 
are suppressed\footnote{We do not consider renormalon ambiguities, instanton 
and other possible non-perturbative effects here.} as 
$(\frac{\Delta_\tau^2}{Q^2})^{\stackrel{(\tau-\tau_{min}) / 2}{\ }}$. 

Our main interest is $\Delta_\tau$ for the next-to-lowest twist operators. 
Twist-4 operators of spin $n$ are related to the $1/Q^2$ power corrections to 
the $n$-th moment of the structure functions:
\bd
\int_0^1 d\x \, \x^{n-2} F_2(\x,Q^2) 
= C^{(2)}_n (Q^2/\mu^2) \, A^{(2)}_n (\mu)
\ed
\be
~~~~~~~~~~~+ C^{(4)}_n (Q^2/\mu^2) \, \frac{A^{(4)}_n (\mu)}{Q^2} \, 
+ \, O(\frac{1}{Q^4}).
\ee
While twist-4 effects are negligible for very large $Q^2$, at energies of a 
few GeV they could be relevant and refine the present QCD predictions. In 
fact, QCD cannot be tested unambiguously unless these contributions are known.

For twist 4, the Wilson coefficients are known at leading order from 
continuum QCD~\cite{ht,js,lww,sv}, but this does not exclude the possibility 
to compute them also non-perturbatively on the lattice~\cite{cf}.
The real physical non-perturbative effects are contained in the 
matrix elements of the operators in the OPE, and the main 
problem is the lack of quantitative knowledge about these matrix elements. 

For the evaluation of these effects a non trivial choice of operator bases for
higher twist operators needs to be made~\cite{ht,js,lww,sv}. The set of all 
possible twist-4 operators is in fact an overcomplete set, and there are 
different ways to eliminate the redundant operators (using the equations of 
motion). A ``canonical'' basis proposed by Jaffe and Soldate~\cite{js} 
involves totally symmetric and traceless operators with no contracted 
derivatives, but even this basis contains more operators than needed, as only 
part of them actually appear in the expansion of the T-product at tree level. 
The relevant basis of non-singlet operators for minimal spin 
is\footnote{Flavor structures are not shown. Purely gluonic operators like 
$F_{\mu\alpha} D_\beta D^\beta F^\alpha_{~\nu}$ belong to the 
singlet sector and appear at higher order in the coupling constant.}:

\vspace{0.3 cm}
\noindent \bt{ll}
$\overline{\psi} \gamma_\mu t^A \psi \, \, \overline{\psi} \gamma_\nu t^A 
\psi~~~~~~$ &$\overline{\psi} \gamma_\mu \gamma_5 t^A \psi \, \,  
\overline{\psi} \gamma_\nu \gamma_5 t^A \psi$ \\
$\overline{\psi} F_{\mu\alpha} \gamma^\alpha \psi~~~~~~$
&$\overline{\psi}\stackrel{\sim}{F}_{\mu\alpha} 
\gamma^\alpha\gamma_5 \psi$ \\
$\overline{\psi} F_{\mu\alpha} F_\nu^{~\alpha} \gamma_\rho \psi~~~~~~$
&$\overline{\psi} \epsilon^{\alpha\beta\lambda\rho}
F_{\mu\alpha} F_{\nu\beta} \gamma_\rho \gamma_5 \psi$ .
\vspace{0.3 cm}
\et
One can also choose a (non-minimal) basis in which operators do not contain
gluon field strengths explicitly, but then contracted derivatives 
appear~\cite{lww}. Most likely, the practical choice of a basis is to be done 
case by case, looking at the complexities of the mixing patterns, at the 
goodness of the signals and at other practical issues. The situation can be 
further complicated by additional mixings due to the lattice.

Sometimes one can simplify the mixing structures. Purely gluonic operators 
can be at once eliminated by including only non-singlet operators from the 
start (as in the basis above), but there also exist particular non-trivial 
combinations of operators that have protection from singlet mixing. As an 
example, the difference between the second moments of longitudinal structure
functions $M_L^{\nu} (2, Q^2) - \frac{18}{5} M_L^{e} (2, Q^2)$ 
is a pure non-singlet quantity also at twist-4 level~\cite{sv}.

Of special importance in the twist-4 case are the 
4-fermion operators. At 1-loop level they do not mix with operators of the 
kind $\overline{\psi} F \psi$, and the mixing matrix is in this respect
triangular. Furthermore, only 4-fermion operators 
can transform as the flavor 27-plet, and this can be exploited to reduce 
strongly the room for renormalization mixing, since the flavor 
27-plet is not present at twist-2 level (where only the singlet and the
octet can contribute). Flavor symmetries can be in general very useful: 
for example the isospin-2 combination of the pion structure 
functions~\cite{morelli},
\vspace{-0.1 cm}
\be
F_{I=2} = F_{\pi^+} + F_{\pi^-} - 2F_{\pi^0},
\ee
gets contributions only from the 27-plet, and therefore cannot mix 
with twist-2 operators.

When mixing with operators of leading twist is forbidden, one can also avoid 
renormalon ambiguities completely. But even in cases where this mixing is 
not prohibited, a computation of the coefficient functions in a 
non-perturbative way~\cite{cf} would get rid of renormalon problems.

\section{PERTURBATIVE RENORMALIZATION}

Renormalization factors are needed to relate the numbers obtained
from lattice simulations to the corresponding physical matrix elements: 
${\cal O}^{cont}(\mu) = Z(\mu a, g(\mu)) \cdot {\cal O}^{latt}(a)$.
In this way one can give ``continuum'' numbers, in the sense that 
a primary result obtained from the lattice can be ``converted'' 
(through these renormalization factors) to its continuum equivalent.

Perturbative lattice renormalization is important by itself and as a 
hint and reference for non-perturbative renormalization studies, especially 
when one has to understand lattice mixings, generally much more intricate
than in the continuum case. Mixing patterns on the lattice are in fact 
more transparent when looked at in perturbation theory.

Here we would like to study the perturbative renormalization
for an important class of twist 4 operators: the 4-quark operators.
We consider the symmetrized operators
\be
{\cal{O}}_{\{\mu\nu\}} = \sum_A \, \overline{\psi} \gamma_{\{\mu} t^A \psi 
\cdot  \overline{\psi} \gamma_{\nu\}} t^A \psi -\mbox{traces}
\ee
\vspace{-0.3 cm}
\bd
{\cal{O}}_{\{\mu\nu\}}^{(5)} = \sum_A \, \overline{\psi} \gamma_{\{\mu} 
\gamma_5 t^A \psi \cdot  \overline{\psi} \gamma_{\nu\}} \gamma_5 t^A \psi
-\mbox{traces} .
\ed
We impose as renormalization conditions that the 1-loop amputated matrix 
elements renormalized at a reference scale $\mu$ are equal to the corresponding
bare tree-level quantities: \pagebreak
\be
\langle q(p) q(p') |{\cal O}(\mu) |q(p) q(p') \rangle = 
\ee
\bd
~~~~~~~~= Z_{{\cal O}} \; \langle q(p) q(p') |{\cal O}(a) |q(p) q(p') \rangle
\Big|^{\rm 1-loop}_{p^2=p'^2=\mu^2}
\ed
\bd
~~~~~~~~= \langle q(p) q(p') |{\cal O}(a)   |q(p) q(p') \rangle 
\Big|^{\rm tree}_{p^2=p'^2=\mu^2}.
\ed
We find it convenient to use forward matrix elements ($p=p'$), the 
renormalization factors being not affected by this choice, and the 
calculations being much simplified.

The computations have been performed using Form codes for the analytic part, 
and Fortran codes for the numerical integration. They have also been checked 
by hand. In principle Fierz transformations could be used (both for color and 
Dirac indices), together with charge conjugation transformations. However, 
they are not needed when suitable sets of diagrams are evaluated together. 
We find that in this way the computations are simpler, and since we can 
integrate numerically each diagram with a good precision in a very short time 
we are not bound to use the 2-fermion results to which the 4-fermion diagrams 
can be reduced by Fierz rearrangements. Furthermore, we avoid any problem 
related to $d$-dimensional Fierz transformations.

From our lattice results we can derive the $Z$ factors to any continuum 
scheme. As an example, we give here the matching factors between the lattice 
and the $\overline{\rm MS}$ scheme at a scale $\mu = \displaystyle{1/a}$. 
With obvious notations, the result for the operator ${\cal{O}}_{\{\mu\nu\}}$ 
(with $\mu\neq\nu$) is:
\bea
&& \hspace{-0.7 cm} Z^{latt\rightarrow\overline{\rm MS}}_{tt}=
(t^A \otimes t^A)\cdot 1 - g_0^2 \cdot \\
&&\cdot \Big\{ 
(t^A \otimes t^A)(-0.001442 + 0.040022 \cdot (1-\alpha)) \nonumber \\
&& +(1 \otimes 1)(-0.023980 + 0.021345 \cdot (1-\alpha)) \Big\}, \nonumber
\eea
where $\alpha$ is the gauge parameter. To complete the mixing matrix we also
need the contribution
\bea
&& \hspace{-0.7 cm} Z^{latt\rightarrow\overline{\rm MS}}_{11}= 
(1 \otimes 1)\cdot 1 - g_0^2 \cdot \\
&& \hspace{-0.2 cm} \cdot 
\Big\{ (1 \otimes 1)(-0.348170) \nonumber \\
&& \hspace{-0.2 cm} +(t_A \otimes t_A) (-0.107911 +0.096053 
\cdot (1-\alpha)) \Big\} . \nonumber
\eea
Similar results are found for the operator ${\cal{O}}_{\{\mu\nu\}}^{(5)}$:
\bea
&& \hspace{-0.7 cm} Z^{latt\rightarrow\overline{\rm MS}}_{tt}= 
(t^A \otimes t^A)\cdot 1 - g_0^2 \cdot \\
&& \hspace{-0.2 cm} \cdot 
\Big\{ (t^A \otimes t^A)(-0.011619 + 0.040022\cdot (1-\alpha)) \nonumber\\
&& \hspace{-0.2 cm} +(1 \otimes 1) (-0.023980 +0.021345 \cdot (1-\alpha)) 
\Big\} \nonumber \\
&& \hspace{-0.7 cm} Z^{latt\rightarrow\overline{\rm MS}}_{11}= 
(1 \otimes 1)\cdot 1 - g_0^2 \cdot \\
&& \hspace{-0.2 cm} \cdot 
\Big\{ (1 \otimes 1)(-0.266750)\nonumber \\
&& \hspace{-0.2 cm} +(t^A \otimes t^A) (-0.107911 +0.096053 \cdot 
(1-\alpha)) \Big\}. \nonumber
\eea
For these calculations we have used a regularization scheme which involves a 
totally anticommuting $\gamma_5$ ($ \{\gamma_5, \gamma_\mu \} =0$), that is 
simple to implement in computer codes. We are considering performing the 
calculations also in the 't Hooft-Veltman scheme.

The calculation of the renormalization factors can also be done 
non-perturbatively~\cite{future}, and it will be interesting to compare the 
perturbative with the non-perturbative results. Other twist-4 operators will 
also be considered, and a choice of basis will have to be made, maybe 
according to the physical situations, so that a reliable estimate of twist-4 
effects for some particular processes can be obtained.


\begin{thebibliography}{99}
\bibitem{ht} S. Gottlieb, Nucl Phys. B139 (1978) 125;
        M. Okawa, Nucl. Phys. B172 (1980) 481 and Nucl. Phys. B187 (1981) 71; 
        R. K. Ellis, W. Furmanski and R. Petronzio, Nucl Phys. B207 (1982) 1
            and Nucl. Phys. B212 (1983) 29. 
\bibitem{js} R. L. Jaffe and M. Soldate, Phys. Lett. 105B (1981) 467 
            and Phys. Rev. D26 (1982) 49.
\bibitem{lww} S. P. Luttrell, S. Wada and B. R. Webber, Nucl. Phys. B188 
            (1981) 219;
        S. P. Luttrell and S. Wada, Nucl. Phys. B197 (1982) 290;
        S. Wada, Nucl. Phys. B202 (1982) 201.
\bibitem{sv} E. V. Shuryak and A. I. Vainshtein, Phys. Lett. 105B (1981) 65 
            and Nucl. Phys. B199 (1982) 451.
\bibitem{cf} D. Petters, these proceedings; S. Caracciolo, these 
proceedings.
\bibitem{morelli} A. Morelli, Nucl. Phys. B392 (1993) 518.
\bibitem{future} S. Capitani et al., in preparation.
\end{thebibliography}
\end{document}